\documentclass[12pt,preprint]{aastex}
\usepackage{epsf}
\newcommand{\simgt}{\lower.5ex\hbox{$\; \buildrel > \over \sim \;$}}
\newcommand{\simlt}{\lower.5ex\hbox{$\; \buildrel < \over \sim \;$}}
\newcommand{\bfsh}{{\bar{\bf s}}}

\begin{document}

\title{Optimal Weighting Scheme in Redshift-space Power Spectrum 
Analysis and 
\\
a Prospect for Measuring the Cosmic Equation of State
}
\author{Kazuhiro Yamamoto}
\affil{
Max-Planck-Institute for Astrophysics,
Karl-Schwarzschild-Str. 1, D-85741 Garching, Germany
\\
Department of Physical Science, 
Hiroshima University, Higashi-Hiroshima, 739-8526, Japan
}
\email{kazuhiro@hiroshima-u.ac.jp}


\begin{abstract}
We develop a useful formula for power spectrum analysis
for high and intermediate redshift galaxy samples, as an
extension of the work by Feldman, Kaiser \& Peacock (1994). 
An optimal weight factor, which minimizes the errors of the
power spectrum estimator, is obtained so that the light-cone 
effect and redshift-space distortions are incorporated. 
Using this formula, we assess the feasibility of the power 
spectrum analysis with the luminous red galaxy (LRG) sample 
in the Sloan Digital Sky Survey as a probe of the equation 
of state of the dark energy.
Fisher matrix analysis shows that the LRG sample can be
sensitive to the equation of state around redshift 
z=0.13. It is also demonstrated that the LRG sample can 
constrain the equation of state with ($1$-sigma) error 
of $10 \%$ level, if other fundamental cosmological 
parameters are well determined independently.
For the useful constraint, we point out the importance 
of modeling the bias taking the luminosity dependence 
into account.
We also discuss the optimized strategy to constrain 
the equation of state using power spectrum analysis.
For a sample with fixed total number of objects,
it is most advantageous to have the sample with the 
mean number density $10^{-4}~h^3{\rm Mpc}^{-3}$ in the 
range of the redshift $0.4 \simlt z\simlt 1$.
\end{abstract}


\maketitle

\newpage
\section{INTRODUCTION}

The clustering of the cosmological objects like galaxies, 
clusters of galaxies and QSOs, is the fundamental to probe  
the Universe because it directly reflects properties of 
dark components and the primordial density fluctuations.
The power spectrum is a simple but very useful tool to 
characterize their spatial distribution. 
Actually, useful constraints on the cosmological 
parameters are obtained from the power spectrum analyses 
of the Two Degree Field (2df) Galaxy Redshift Survey 
(Percival et~al. 2001) and the 2df QSO Redshift Survey 
(Hoyle et~al. 2002).
In such power spectrum analyses, many authors base
their methods on the seminal paper by Feldman, Kaiser 
\& Peacock (1994, Hereafter FKP).

For redshift surveys such as the 2df survey and the Sloan 
Digital Sky Survey (SDSS), however, several observational 
effects on the power spectrum analysis can be very important. 
Because cosmological observations are feasible only on the
light-cone hypersurface defined by the current observer,
the effect of the redshift evolution of the 
luminosity function, the clustering amplitude, and the 
bias, contaminates the observational data. We call this
the light-cone effect (Matarrese et al. 1997; Matsubara, 
Suto \& Szapudi 1997; de Laix \& Starkman 1998; 
Yamamoto \& Suto 1999).
On the other hand it is well known that the distribution
of the sources in redshift sapce is different from that 
in the real space due to the redshift-space distortions.
The linear redshift distortion is the effect of the bulk 
motion of the sources within the linear theory of 
density perturbation (Kaiser 1987; Hamilton 1998 and references 
therein). 
The finger of Got effect is the redshift distortion due to
the random motion of the sources in the nonlinear regime
(Mo, Jing \& Boerner 1997; Magira, Jing \& Suto 2000). 
The geometric distortion is the effect caused by a choice of 
the distance-redshift relation to plot a map of the sources 
(Ballinger, Peacock \& Heavens 1996; Matsubara \& Suto 1996). 

The LRG sample of the SDSS spectroscopic survey will 
provide a sample of $10^5$ intrinsically luminous early-type
galaxies to $z\sim0.5$ (Eisenstein et~al. 2001).
In the analysis of clustering statistics of such a sample, 
the light-cone effect and the redshift-space distortions 
can be substantial. 
Fortunately these observational effects have been well 
investigated and we can model the power spectrum incorporating 
them (see section 3.1).
The primary purpose of the present paper is to extend 
the formulas in FKP in order to incorporate these 
observational effects and to derive a generalized 
expression of the optimal weight factor in the power 
spectrum analysis (cf. Tegmark 1995). 

The results of the 2df galaxy survey (Peacock et~al. 2001), 
weak lensing surveys (Refregier et~al. 2002; Bacon et~al. 2002; 
and references therein), 
cosmic microwave background anisotropy measurements (e.g., Ruhl et~al. 2002) 
and type Ia supernovae measurements (Perlmutter et~al. 1999;
Riess et al. 1998), 
support the concordance model (Wang et~al. 2000):
A spatially flat universe dominated by the dark energy 
component and, with respect to structure formation, 
cold dark matter with the primordial density fluctuations 
predicted in the inflationary scenario.
The very recent result by the Wilkinson Microwave Anisotropy 
Probe (WMAP) strongly supports the concordance model (Spergel et~al. 2003).
Now the mystery of the dark energy component has become one 
of the most important issues in cosmology, which has led to
recent activity in investigating the dark energy (See e.g. 
Peebles \& Ratra 2003, for a recent review, and references therein). 
The dark energy can be characterized by its equation of 
state $w_X=p_X/\rho_X$, where $p_X$ is the pressure and
$\rho_X$ is the energy density. 
For the cosmological constant, $w_X=-1$.
However, if the dark energy originates from the vacuum 
energy of a variable scalar field, 
like the quintessence model, $w_X$ can take values
$w_X>-1$, and can in general be a function of redshift.
Thus constraints
on the equation of state is quite important in considering
the origin of the dark energy. Then various strategies for 
probing the equation of state have been investigated (e.g. 
Newman \& Davis 2000, Saini, et~al. 2000, Wang, et~al. 2000, 
Chiba \& Nakamura 2000, Huterer \& Turner 2001 , 
Yamamoto \& Nishioka 2001, Kujat et~al. 2002, and references therein). 

Second purpose of this paper is to assess the feasibility of 
measuring the equation of state using the power spectrum 
analysis.
\footnote{After completing this work, a similar investigation 
by Matsubara \& Szalay (2003) has been announced.}
Recently, Matsubara \& Szalay have discussed the usefulness 
of the LRG sample in SDSS for measuring the cosmological 
parameters (2002). Their method is based on maximum 
likelyhood analysis in redshift space. 
They demonstrated the usefulness of various SDSS samples
to constrain cosmological parameters by estimating the 
Fisher matrix element. Motivated by their work, 
we assess the feasibility of the method with the power spectrum 
analysis of the LRG sample in SDSS to probe the equation 
of state based on the Fisher matrix formalism for power spectrum 
analysis (Tegmark 1997, Tegmark et~al. 1998a).
The advantage of our method is that the formulae are simple 
and analytic, which allows us to evaluate the formulae easily 
and to understand meaning of results clearly.

This paper is organized as follows: In section 2, we derive 
formulae for the power spectrum analysis taking the 
light-cone effect and the redshift-space distortions into account. 
In section 3, the Fisher matrix element is evaluated 
using an approximate method. 
There we focus on the matrix elements relevant to a measurement
of dark energy, especially the equation of state. 
In section 4, we discuss optimizing the strategy, for a sample 
with a fixed total number of objects, using  the power spectrum 
to probe the equation of state.
Section 5 is devoted to summary and conclusions.
Throughout this paper we use a system of units in which the
velocity of light $c$ equals $1$.

\def\bfs{{\bf s}}
\def\bfk{{\bf k}}
\def\calF{{\cal F}}
\def\calP{{\cal P}}
\def\DeltaV{{V}}
\def\bfF{{F}}

\section{FORMALISM}
In this section we present an optimal weighting scheme 
of the power spectrum analysis for redshift surveys 
taking the redshift-space distortions 
and the light-cone effect into account.
We obtain the generalized optimal weight factor, which 
minimizes errors of a power spectrum estimator, as a
simple extension of the scheme developed by FKP. 
On the basis of this result, we derive the expression
of the Fisher matrix element in subsection 2.2.

In a redshift survey, the redshift $z$ is the indicator
of the distance, therefore, we need to {\it assume} a 
distance-redshift relation to plot a map of objects.
For this distance, in the present paper, 
we adopt the following expression, which is the 
comoving distance in spatially flat universes
\begin{equation}
  s(z)={1\over H_0}\int_0^z{dz'\over
  \sqrt{\Omega_0(1+z')^3+1-\Omega_0}},
\label{defs}
\end{equation}
where $H_0=100h{\rm km/s/Mpc}$ is the Hubble parameter 
and we fix $\Omega_0=0.3$. Our formula presented below 
is general,  and it does not depend on this choice of 
$s(z)$.

\subsection{Optimal Weight Factor}
The derivation of the optimal weight factor in this subsection
is essentially same as that of FKP. Some parts of equations
here can be found in FKP, which we have not omitted 
for being self-contained of the present paper.
We denote the number density field of sources (real catalog) 
by $n_g(\bfs)$, where $\bfs(=s(z)\gamma)$ is the three 
dimensional coordinate in the (cosmological) redshift space 
and $\gamma$ is the unit directional vector. 
We use $\bar n(\bfs)$ to denote the expected mean number density.
We define the fluctuation field to be
\begin{eqnarray}
  F(\bfs)=n_g(\bfs)-\alpha n_s(\bfs),
\label{defF}
\end{eqnarray}
where $n_g(\bfs)=\sum_i\delta(\bfs-\bfs_i)$, 
with $\bfs_i$ being the location of the $i-$th object, 
similarly $n_s(\bfs)$ is the density of a synthetic catalog 
which has mean number density $1/\alpha$ times that of the 
real catalog. For $n_g(\bfs)$ and $n_s(\bfs)$, following FKP, 
we assume
\begin{eqnarray}
  &&\langle n_g(\bfs_1)n_g(\bfs_2)\rangle=
  \bar n(\bfs_1)\bar n(\bfs_2)(1+\xi(\bfs_1,\bfs_2))
  +\bar n(\bfs_1)\delta(\bfs_1-\bfs_2),
\label{nnA}
\\ 
 &&\langle n_s(\bfs_1)n_s(\bfs_2)\rangle=
  \alpha^{-2}\bar n(\bfs_1)\bar n(\bfs_2)
  +\alpha^{-1}\bar n(\bfs_1)\delta(\bfs_1-\bfs_2),
\label{nnB}
\\ 
 &&\langle n_g(\bfs_1)n_s(\bfs_2)\rangle=
  \alpha^{-1}\bar n(\bfs_1)\bar n(\bfs_2),
\label{nnC}
\end{eqnarray}
where $\xi(\bfs_1,\bfs_2)$ denotes the correlation function.

We define the Fourier coefficient of $F(\bfs)$, with a 
conventional weight factor, to be
\begin{eqnarray}
  \calF(\bfk)={\int d\bfs  w(\bfs,\bfk) F(\bfs) e^{i\bfk\cdot \bfs}
  \over 
  [\int d\bfs \bar n(\bfs)^2 w(\bfs,\bfk)^2]^{1/2}},
\label{defcalF}
\end{eqnarray}
where $w(\bfs,\bfk)$ is the weight function which can be adjusted 
to optimize the power spectrum estimator below.
Then the expectation value of the square $|\calF(\bfk)|^2$ is 
written, 
\begin{eqnarray}
  &&\langle |\calF(\bfk)|^2\rangle={
 \int d\bfs_1 \int d\bfs_2
 w(\bfs_1,\bfk)w(\bfs_2,\bfk)
 \langle F(\bfs_1) F(\bfs_2)\rangle 
  e^{i\bfk\cdot (\bfs_1-\bfs_2)} \over
  \int d\bfs \bar n(\bfs)^2 w(\bfs,\bfk)^2}.
\nonumber
\\
\end{eqnarray}
Using equations (\ref{nnA})-(\ref{nnC}), we have
\begin{eqnarray}
 &&\langle F(\bfs_1) F(\bfs_2)\rangle 
  =\bar n(\bfs_1)\bar n(\bfs_2)\xi(\bfs_1,\bfs_2)
  +(1+\alpha)\bar n(\bfs_1)\delta(\bfs_1-\bfs_2).
\end{eqnarray}
Under the distant observer approximation, $|\bfs_1-\bfs_2|\ll|\bfs_1|,
|\bfs_2|$,  the 
correlation function can be expressed as 
\begin{eqnarray}
  \xi(\bfs_1,\bfs_2)=\int{d\tilde\bfk\over (2\pi)^3}
  P(\tilde \bfk,|\bfsh|)
  e^{-i\tilde\bfk\cdot(\bfs_1-\bfs_2)},
\end{eqnarray}
where $P(\tilde \bfk,|\bfsh|)$ is the redshift-space 
power spectrum at the distance $|\bfsh|$, where we defined 
$\bfsh=(\bfs_1+\bfs_2)/2$. We also adopt the approximation 
\begin{eqnarray}
  w(\bfs_1,\bfk)w(\bfs_2,\bfk)\bar n(\bfs_1)\bar n(\bfs_2)
  \simeq w(\bfsh,\bfk)^2\bar n(\bfsh)^2,
\end{eqnarray}
which yields
\begin{eqnarray}
  &&\langle |\calF(\bfk)|^2\rangle=
  {\int d\bfsh \bar n(\bfsh)^2 w(\bfsh,\bfk)^2 P(\bfk,|\bfsh|)
  +(1+\alpha)\int d\bfs \bar n(\bfs) w(\bfs,\bfk)^2 \over
  \int d\bfs \bar n(\bfs)^2 w(\bfs,\bfk)^2 }.
\label{calFF}
\end{eqnarray}
The last term of the right hand side of equation (\ref{calFF})
corresponds to the shot-noise, therefore the 
estimator of the power spectrum should be defined as
\begin{eqnarray}
  \calP(\bfk)=|\calF(\bfk)|^2-P_{\rm shot}(\bfk),
\label{defcalP}
\end{eqnarray}
where $P_{\rm shot}$ is defined
\begin{eqnarray}
 P_{\rm shot}(\bfk)={(1+\alpha)\int d\bfs \bar n(\bfs) w(\bfs,\bfk)^2
  \over \int d\bfs \bar n(\bfs)^2 w(\bfs,\bfk)^2}.
\end{eqnarray}
The angular average of $\calP(\bfk)$ over a thin shell 
of the radius $k(=|\bfk|)$ in the Fourier space gives 
the estimator of the angular averaged power spectrum
\begin{eqnarray}
 \calP_0(k)={1\over \DeltaV_{k} }
  \int_{\DeltaV_{k}} d\bfk \calP(\bfk),
\label{defcalPzero}
\end{eqnarray}
where $\DeltaV_{k}$ denotes the volume of the shell. 
Thus the expectation value of $\calP_0(k)$ gives
the angular averaged power spectrum incorporating the
light-cone effect 
\begin{eqnarray}
  \langle\calP_0(k)\rangle
  ={\displaystyle{\int d\bfs \bar n(\bfs)^2 w(\bfs,\bfk)^2 P_0(k,|\bfs|) }
  \over\displaystyle{{\int d\bfs \bar n(\bfs)^2 w(\bfs,\bfk)^2 }}},
\label{lightPk}
\end{eqnarray}
where
\begin{eqnarray}
P_0(k,|\bfs|)=
{1 \over \DeltaV_{k} }\int_{\DeltaV_{k}} d{\bfk}  P(\bfk,|\bfs|).
\end{eqnarray}

Next we consider the variance of $\calP(\bfk)$, defined by
\begin{eqnarray}
  \langle \Delta \calP(\bfk) \Delta \calP(\bfk') \rangle
&=&
  \langle [\calP(\bfk)-\langle \calP(\bfk)\rangle]
          [\calP(\bfk')-\langle \calP(\bfk')\rangle]\rangle
\nonumber\\
&=&  \langle \calP(\bfk)\calP(\bfk') \rangle -
  \langle \calP(\bfk)\rangle
  \langle \calP(\bfk') \rangle,
\label{defPPA}
\end{eqnarray}
which is rephrased, by using (\ref{defcalP}),
\begin{eqnarray}
  \langle \Delta \calP(\bfk) \Delta \calP(\bfk') \rangle =
  \langle |\calF(\bfk)|^2|\calF(\bfk')|^2 \rangle 
  -\langle |\calF(\bfk)|^2 \rangle
     \langle |\calF(\bfk')|^2\rangle.
\label{defPPB}
\end{eqnarray}
Substituting (\ref{defcalF}) into equation (\ref{defPPB}), we have
\begin{eqnarray}
  &&\langle \Delta \calP(\bfk) \Delta \calP(\bfk') \rangle=
\nonumber
\\
  && {2 \prod_{i=1}^2 [\int d\bfs_i w(\bfs_i,\bfk)]
       \prod_{j=3}^4 [\int d\bfs_j w(\bfs_j,\bfk')]
  \langle F(\bfs_1)F(\bfs_3)\rangle
   \langle F(\bfs_2)F(\bfs_4)\rangle
   e^{i\bfk\cdot(\bfs_1-\bfs_2)}e^{-i\bfk'\cdot(\bfs_3-\bfs_4)}
\over [\int d\bfs \bar n(\bfs)^2 w(\bfs,\bfk)^2]
      [\int d\bfs'\bar n(\bfs')^2w(\bfs',\bfk')^2]},
\nonumber
\\
\label{defPPC}
\end{eqnarray}
where we assumed the following relation, which is exact 
when $F(\bfs)$ follows the statistics of 
the Gaussian random field, 
\begin{eqnarray}
   \langle F(\bfs_1)F(\bfs_2)F(\bfs_3)F(\bfs_4)\rangle &=&
   \langle F(\bfs_1)F(\bfs_2)\rangle
   \langle F(\bfs_3)F(\bfs_4)\rangle
\nonumber
\\
   &+&\langle F(\bfs_1)F(\bfs_3)\rangle
   \langle F(\bfs_2)F(\bfs_4)\rangle
\nonumber
\\
  &+&\langle F(\bfs_1)F(\bfs_4)\rangle
   \langle F(\bfs_3)F(\bfs_2)\rangle.
\end{eqnarray}
In a similar way to derive equation (\ref{calFF}), 
using the distant observer approximation, 
one can have the relation 
\begin{eqnarray}
  &&\int d\bfs_1\int d\bfs_3 w(\bfs_1,\bfk) w(\bfs_3,\bfk')
  \langle F(\bfs_1)F(\bfs_3)\rangle
  e^{i\bfk\cdot\bfs_1-i\bfk'\cdot\bfs_3}
\nonumber
\\
  &&\hspace{2cm}\simeq \int d \bfsh \bar n(\bfsh)^2 
  w(\bfsh,\bfk)w(\bfsh,\bfk')
  e^{i\bfsh\cdot(\bfk-\bfk')} 
  \Bigl[
  P\Bigl({\bfk+\bfk'\over 2},|\bfsh|\Bigr)
  +{\alpha+1 \over \bar n(\bfsh)}\Bigr].
\end{eqnarray}
Then, with a repeated use of the distant observer approximation,
we have 
\begin{eqnarray}
  &&\langle \Delta \calP(\bfk) \Delta \calP(\bfk') \rangle\simeq
  {2\int d\bar\bfs \bar n(\bar\bfs)^4 w(\bar\bfs,\bfk)^4\left(
  P(\bfk, \bar\bfs)+{(1+\alpha)/ \bar n(\bar \bfs)}\right)^2(2\pi)^3\delta^{(3)}(\bfk-\bfk')
\over [\int d\bfs \bar n(\bfs)^2 w(\bfs,\bfk)^2]^2}.
\nonumber
\\
\label{defPPGG}
\end{eqnarray}

From (\ref{defcalPzero}), the variance of $\calP_0(k)$ 
is obtained by 
\begin{eqnarray}
  \langle \Delta \calP_0(k)^2 \rangle &\equiv&
  \langle [\calP_0(k)-  \langle\calP_0(k)\rangle]^2 \rangle
\nonumber
\\
  &=&  {1\over \DeltaV_{k}^2}  
  \int_{\DeltaV_k} d\bfk \int_{\DeltaV_k} d\bfk'
  \langle \Delta \calP(\bfk) \Delta \calP(\bfk') \rangle,
\label{defPPE}
\end{eqnarray}
which reduces to
\begin{eqnarray}
   && \langle \Delta \calP_0(k)^2 \rangle=
  2{(2\pi)^3\over \DeltaV_{k}}
  {\int d\bfsh \bar n(\bfsh)^4 w(\bfsh,\bfk)^4
  Q^2(k,\bfsh) \over
  [\int d\bfs \bar n(\bfs)^2 w(\bfs,\bfk)^2]^{2}},
\label{defPPF}
\end{eqnarray}
where we defined
\begin{eqnarray}
  Q^2(k,\bfsh) = {1\over \DeltaV_{k}}
  \int_{\DeltaV_k} d\bfk   \Bigl[
  P\Bigl({\bfk},|\bfsh|\Bigr)
  +{\alpha+1 \over \bar n(\bfsh)}\Bigr]^2.
\label{defPPG}
\end{eqnarray}
The optimal weight factor $w(\bfs,\bfk)$ which minimizes  
$\langle\Delta \calP_0(k)^2\rangle$ is found 
from the stationary solution against the
variation $w\rightarrow w+\delta w$, to be 
\begin{eqnarray}
  w(\bfs,k)={1\over \bar n(\bfs) Q(k,\bfs) }.
\label{defPPH}
\end{eqnarray}
This generalized formula incorporates
the redshift distortions. In general, the
redshift space power spectrum $P(\bfk,s)$ is written,
in terms of the multipole expansion,
\begin{eqnarray}
  P(\bfk,s)=\sum_{l=0,2,\cdots} P_l(k,s){\cal L}_l(\mu),
\end{eqnarray}
where $\mu$ is the directional cosine between the line
of sight $\gamma$ and the wave number vector,
$\mu=\cos(\gamma\cdot \bfk/k)$, ${\cal L}_l(x)$ is 
the Legendre polynomial of the $l$-th order, and 
$P_l(k,s)$ are the expansion coefficients.
Then (\ref{defPPG}) reduces to
\begin{eqnarray}
  Q^2(k,\bfsh) = \biggl(P_0(k,|\bfsh|)+{1\over \bar n(\bfsh)}\biggr)^2
                 +P_2(k,|\bfsh|)^2+P_4(k,|\bfsh|)^2+\cdots,
\label{defPPI}
\end{eqnarray}
where we considered the limit $\alpha\ll1$. 
In practice, the contribution of higher 
moments $l\ge2$ is not large (cf. Yamamoto 2003). 
Neglecting this higher moments and the redshift
dependence of the power spectrum, (\ref{defPPH})
reduces to the similar expression by FKP (cf. Peebles 1980),
\begin{eqnarray}
  w(\bfs,k)\simeq {1\over 1+\bar n(\bfs) P_0(k,|\bfs|)},
\label{apw}
\end{eqnarray}
then, from equation (\ref{defPPF}), we have 
\begin{eqnarray}
   && \langle \Delta \calP_0(k)^2 \rangle=
  2{(2\pi)^3\over \DeltaV_{k}}
  \biggl[\int d\bfs {\bar n(\bfs)^2\over
  (1+\bar n(\bfs) P_0(k,|\bfs|))^2 }\biggr]^{-1}.
\label{defPPM}
\end{eqnarray}

Here we mention a technical problem of the above optimal weight factor.
Namely, this weight factor contains the power spectrum at each 
redshift, i.e. $P_0(k,s[z])$. It might be difficult to evaluate 
it with a sufficient accuracy from a observational data set. 
A possible alternation is to adopt the weight factor:
\begin{eqnarray}
  w(\bfs,k)= {1\over 1+\bar n(\bfs) \langle \calP_0(k)\rangle}.
\label{apwB}
\end{eqnarray}
This modified weight factor does not minimize the errors, however, 
it might be useful in a realistic situation of data analysis. 
The weight factor contains $\langle \calP_0(k)\rangle$, 
then it must be solved with equation (\ref{lightPk}) as 
a coupled system. However, we can easily solve the 
coupled equation numerically by iterations.
In this case the amplitude of the covariance matrix 
$\langle \Delta \calP_0(k)^2 \rangle$ 
is different from the expression (\ref{defPPM}).
It is obtained by substituting (\ref{apwB}) 
into (\ref{defPPF}), but the expression is rather 
complicated than (\ref{defPPM}).
As we show below, the use of the modified weight factor 
alters the expression of the Fisher matrix element. However, 
the difference is not significant. For example, the use 
of the modified weight factor increases the $1-\sigma$
error of the equation of state $\Delta\bar w$ (see below) 
by 10 \% level. Then our conclusion does not depend on 
the choice of the above weight factors.

\newcommand{\cpara}{c_{\scriptscriptstyle \|}}
\newcommand{\cperp}{c_{\scriptscriptstyle \bot}}
\newcommand{\qpara}{q_{\scriptscriptstyle \|}}
\newcommand{\qperp}{q_{\scriptscriptstyle \bot}}

\subsection{Fisher Matrix Element}
In order to estimate the accuracy to which we can
constrain cosmological parameters with a measurement
of the power spectrum, we employ the Fisher matrix
approach. With the Fisher-matrix analysis, one can
estimate the best statistical errors on parameters
from a given data set (Kendall \& Stuart 1969). 
For this reason, this approach 
is widely used to estimate how accurately cosmological
parameters are determined from large surveys, such 
as the large scale structure of galaxies, cosmic 
microwave background anisotropy measurement and 
supernova data sets. (See e.g. Tegmark, Taylor, \& 
Heavens 1997, Jungman et~al. 1996a, 1996b, 
Zaroubi et~al. 1995, Fisher et~al. 1995, 
Tegmark et~al. 1998b).
In general, the Fisher matrix is defined by
\begin{eqnarray}
  {\bfF}_{ij}=-\Bigl\langle {\partial^2 \ln L
  \over \partial\theta_i \partial\theta_j}
  \Big\rangle,
\end{eqnarray}
where $L$ is the probability distribution function of a data 
set, given model parameters $\theta_i$. For a measurement of 
galaxy power spectrum, the Fisher matrix is presented using 
a suitable approximation in the references 
(Tegmark 1997, Tegmark et~al. 1998a). 
Here we briefly review a derivation of the Fisher matrix:
For simplicity, we adopt the approximation of the Gaussian
probability distribution function for $\Delta \calP(\bfk)$,
\begin{eqnarray}
  L\propto \exp\left[ -{1\over 2} \int d\bfk \int d\bfk' \Delta \calP(\bfk)
  C(\bfk,\bfk') \Delta \calP(\bfk')\right],
\end{eqnarray}
where $C(\bfk,\bfk')$ is the inverse matrix of the covariance 
matrix $\langle \Delta \calP(\bfk) \Delta \calP(\bfk') \rangle$. 
Then, using (\ref{defPPG}), we have 
\begin{eqnarray}
  {\bfF}_{ij}\simeq {1\over 4\pi^2} 
  \int _{k_{\rm min}}^{k_{\rm max}} 
  \kappa(k)
  {\partial \langle\calP_0(k)\rangle\over \partial \theta_i}
  {\partial \langle\calP_0(k)\rangle\over \partial \theta_j}
  k^3 d\ln k,
\label{kint}
\end{eqnarray}
where $\kappa(k)$ is
\begin{eqnarray}
  \kappa(k)^{-1}=
    {\int d\bar\bfs \bar n(\bar\bfs)^4 w(\bar\bfs,k)^4\left(
  P_0(k, \bar\bfs)+{1/ \bar n(\bar \bfs)}\right)^2
  \over [\int d\bfs \bar n(\bfs)^2 w(\bfs,k)^2]^2},
\end{eqnarray}
and $\langle\calP_0(k)\rangle$ is the expectation value 
of the power spectrum, i.e. expression (\ref{lightPk}).
The Fisher matrix depends on the weight factor, and we 
have
\begin{eqnarray}
  \kappa(k)=\Delta\Omega \int ds s^2 {\bar n (s)^2 \over 
  (1+\bar n(s) P_0(k,s))^2 }
\end{eqnarray}
for the optimal weight factor (\ref{apw}), and
\begin{eqnarray}
  \kappa(k)={\Delta\Omega[\int ds s^2 \bar n(s)^2 (1+\bar n(s)\langle\calP_0(k)\rangle)^{-2}]^2
  \over \int ds s^2 \bar n(s)^4 (P_0(k,s)+1/\bar n(s))^2 
 (1+\bar n(s)\langle\calP_0(k)\rangle)^{-4} }
\end{eqnarray}
for the modified weight factor (\ref{apwB}), respectively,
where we assumed that the mean number density is a 
function of the distance $s(=|\bfs|)$, and 
$\Delta\Omega$ is the survey area.
The result is consistent with the previous work:
In the limit that the linear redshift distortion, the geometric distortion, 
and the light-cone effect are switched off, the above results reproduce 
the previous work (e.g., Tegmark 1997).
By using the Bayes theorem, the probability distribution in the parameter
space can be written 
\begin{eqnarray}
  p(\theta_i)\propto \exp\biggl[-{1\over 2}\sum_{ij}
  (\theta_i-\theta_i^{\rm tr})F_{ij}
  (\theta_j-\theta_j^{\rm tr})\biggr],
\label{probp}
\end{eqnarray}
where we have assumed that errors in the target model 
parameters $\theta_i^{\rm tr}$ are small.
Thus the Fisher matrix gives the uncertainties in the 
parameter spaces, which are described by a Gaussian
distribution function around $\theta_i^{\rm tr}$. 

\section{CONSTRAINTS}
In this section we investigate a prospect of the power spectrum 
analysis of the LRG sample in the SDSS. As pointed 
out by Matsubara \& Szalay (2002), the LRG sample can be a useful 
tool to constrain the cosmological parameters. 
Here we assess the potential of the LRG sample to 
constrain the equation of state of the dark energy. 
Because the LRG sample is distributed
out to a redshift $\sim 0.5$, the geometric distortion is substantial.
This is the reason why the power spectrum analysis 
of the LRG sample can constrain the equation of 
state $w_X$ even if the original matter power spectrum 
(or the transfer function) does not depend on $w_X$. 
In comparison to the LRG sample, the SDSS quasars are distributed 
out to $z\simeq 5$, however, the spatial distribution is very sparse. 
Then the constraint from the quasar sample 
will not be very tight due to the large shot-noise 
contribution (Matsubara \& Szalay 2002, Yamamoto 2002).

The clustering properties of the LRG sample have not
been analyzed so far, then we here assume simple linear 
bias models for the clustering. (See equation (\ref{bz}) 
and section 3.3 for a luminosity dependent bias model).
To evaluate the Fisher matrix element, the number density 
of the sample $\bar n(z)$ is an important factor. 
In the present paper, we adopt the LRG sample with the 
comoving number density in the reference by Eisenstein 
et~al. (2001, Figure 12 in their paper), as a function 
of the redshift in the range $0.2 \leq  z \leq 0.55$.  
The typical number density of LRGs is $\bar n\simeq 10^{-4} 
~[h^{-1}{\rm Mpc}]^{-3}$. The peak value of the power 
spectrum is $10^{5}~[h^{-1}{\rm Mpc}]^3$. 
Thus $\bar n(z) P_0(k,z)$ does not exceed $10$. 

\subsection{Modeling the Power Spectrum}

We restrict ourselves to a spatially flat universe for 
modeling the cosmology. This assumption can be
justified by the inflationary universe scenario and 
the recent results of the cosmic microwave anisotropy 
measurements. We consider a cosmological 
model with the dark energy component with the 
equation of state which is variable in time.
We assume that the time variation of the equation of 
state $w_X(z)=p_X/\rho_X$ is slow. The equation of 
motion, $d(\rho_X V(z))+p_XdV(z)=0$ with a volume 
$V(z)\propto (1+z)^{-3}$, yields
\begin{eqnarray}
  {\rho_X(z) \over \rho_X(z=0)}= (1+z)^3\exp\biggl[3\int_0^z
  {w_X(z')\over 1+z'} dz'\biggr]\equiv f_X(z).
\end{eqnarray}
With the parameters $\bar w$ and $\nu$, we assume 
$w_X$ can be parametrized by 
\begin{equation}
  w_X(z)=\bar w \biggl({1+z\over 1+z_*}\biggr)^\nu,
\label{wz}
\end{equation}
where $z_*$ is a constant, then we have
\begin{eqnarray}
  f_X(z)=(1+z)^3\exp\biggl[{3{\bar w}\over (1+z_*)^\nu}\biggl(
  {(1+z)^\nu-1\over \nu} \biggr)\biggr].
\end{eqnarray}
Denoting the matter density parameter 
by $\Omega_m$, the comoving distance is 
\begin{equation}
  r(z)={1\over H_0}\int_0^z{dz'\over
  \sqrt{\Omega_m(1+z')^3+(1-\Omega_m)f_X(z')}}.
\label{defr}
\end{equation}

We can model the power spectrum in redshift space $s(z)$ as
(Ballinger, Peacock, \& Heavens 1996)
\begin{eqnarray}
  &&P_0(k,s(z))={1\over \cpara(z)\cperp(z)^2}
  \int_0^1d\mu
  P_{\rm gal}\Bigl(\qpara\rightarrow{k\mu\over\cpara},
  \qperp\rightarrow{k\sqrt{1-\mu^2}\over\cperp},z\Bigr),
\end{eqnarray}
where $P_{\rm gal}(\qpara,\qperp,z)$ is the galaxy power
spectrum,  $\qpara$ and $\qperp$ are the wave number
components parallel and perpendicular to the line-of-sight
direction in real space with $r(z)$, and we define
\begin{eqnarray}
&&\cpara(z)={dr(z)\over ds(z)}
\hspace{.5cm}{\rm and}\hspace{.5cm}
\cperp(z)={ r(z)\over  s(z)}.
\end{eqnarray}
The geometric distortion is 
described by scaling the wave number by $\cpara(z)$ and $\cperp(z)$.
Then, even if the original matter power spectrum or the transfer function 
does not depend on the parameter of the dark energy $w_X$, 
the power spectrum observed in redshift-space can depend on $w_X$.
Thus we can test the nature of dark energy. 
This idea traces back to the geometric test pointed out by 
Alcock \& Paczynski (1979).
We model the power spectrum in the distribution of galaxy by
\begin{eqnarray}
  P_{\rm gal}(\qpara,\qperp,z)=
  \biggl(1+{f(z)\over b(z)}{\qpara^2\over q^2}\biggr)^2
  b(z)^2 P_{\rm mass}(q,z),
\label{PQSO}
\end{eqnarray}
where $q^2=\qpara^2+\qperp^2$, $b(z)$ is a scale independent
bias factor, $P_{\rm mass}(q,z)$ is the CDM mass power 
spectrum, and we defined $f(z)=d\ln D(z)/d\ln a(z)$ with
the linear growth rate $D(z)$ and the scale factor $a(z)$.
The term in proportion to $f(z)$ describes the linear 
distortion (Kaiser 1987). 
Following the work by Matsubara \& Szalay (2002), we work within the linear 
theory of density perturbation, adopting the fitting 
formula of the transfer function by Eisenstein \& Hu (1998).
Here we consider a simple bias model parametrized by, 
\begin{eqnarray}
  b(z)=1+(b_0-1){1\over D(z)^p},
\label{bz}
\end{eqnarray}
where $b_0$ and $p$ are constants. 
For $f(z)$ we adopt the fitting formula
developed by Wang and Steinhardt (1998)
\begin{eqnarray}
 f(z)\simeq \Omega(z)^{\alpha(w_X)},
\end{eqnarray}
with 
\begin{eqnarray}
  &&\alpha(w_X)={3\over 5-w_X/(1-w_X)}+{3\over 125}
  {(1-w_X)(1-3w_X/2)\over (1-6w_X/5)^3}(1-\Omega(z)),
\\ 
  &&\Omega(z)={\Omega_m\over \Omega_m
  +(1-\Omega_m)(1+z)^{-3}f_X(z)}.
\end{eqnarray}
In the present paper, we work within the linear theory of 
density perturbations and we do not consider the nonlinear 
and the finger of Got effects. 
The nonlinear effect is substantial on small scales in the 
nonlinear regime (Mo, Jing \& Boerner 1997; 
Magira, Jing \& Suto 2000). The inclusion of the nonlinear
effect will not alter our result significantly, however, 
any uncertainty of modeling the nonlinear effect reduces
the capability to determine cosmological parameters 
precisely (cf. Watkins et~al. 2002).

\subsection{Constraints on $w_X$}

In our analysis we focus on the sensitivity of the power
spectrum analysis on the nature of the dark energy, 
then we consider the parameters $(\Omega_m,\bar w,\nu,b_0,p)$.
We assume that other cosmological parameters such 
as the baryon density $\Omega_b$, $\sigma_8$ and the index of the 
initial density power spectrum $n$ are well determined. 
We might expect to obtain such information in the near 
future from other observations of the cosmic 
microwave background anisotropies and the large 
scale structure of the main galaxy sample in SDSS, etc. 
Actually the WMAP result have demonstrated the cosmological
parameters can be determined from the cosmic microwave 
background anisotropies. In the present paper, we take
$\Omega_m=0.28$,
$\Omega_bh^2=0.024$, $h=0.7$, $\sigma_8=0.9$ and $n=1$
(Spergel et al. 2003).
On the other hand, the bias is an annoying problem, 
though the physical mechanism has been extensively 
investigated (e.g. Mo \& White 1996).
Recent investigation reports that the bias depends on 
the luminosity and galaxy type (Norberg et~al. 
2001, 2002). 
Thus the bias depends on sample, and it should
be determined from the same data simultaneously. 
In our investigation, we here consider the 
marginalized probability function integrating over 
the bias parameters $b_0$ and $p$ in (\ref{probp}). 

Figure 1 shows the contour of the marginalized probability 
function on $\Omega_m-\bar w$ space, which is obtained by 
integrating (\ref{probp}) with respect to $(\nu,~b_0,~p)$.
Here, we adopted the target parameters 
$(\theta_i^{\rm tr})$ as follows: $\Omega_m=0.28$, $\bar w=-1$, 
$\nu=0$, $b_0=1.5$ and $p=1$. The target parameter
of the equation of state corresponds to the cosmological 
constant. The dependence of the bias modeling is discussed
in more detail in section 3.3.
In (\ref{wz}), we here set $z_*=0.13$. We also assume the 
survey area $10^4~{\rm deg}^2$, which corresponds to the
complete SDSS sample.\footnote{
The dependence on the survey area will be discussed in the
final section.} We adopted the range of 
the integration in (\ref{kint}) being  
$k_{\rm min}=0.001~h{\rm Mpc}^{-1}$ and 
$k_{\rm max}=1~h{\rm Mpc}^{-1}$.
Our results do not depend on $k_{\rm min}$ 
as long as $k_{\rm min}\simlt 0.01~h{\rm Mpc}^{-1}$,
while depending slightly on $k_{\rm max}$, but our
conclusions are not significantly altered as long 
as $k_{\rm max}\simgt 0.3~h{\rm Mpc}^{-1}$.
It is clear from Figure 1 that the power spectrum 
is more sensitive to $\Omega_m$ than $\bar w$. 
In our modeling the transfer function depends on 
$\Omega_m$, but not on $w_X$. The effect of $w_X$ 
on the power spectrum comes from the geometric
distortion. The weaker dependence of the power 
spectrum on $\bar w$ than $\Omega_m$ originates from this fact.

Figure 2 shows the contour of the marginalized probability
function on $\bar w-\nu$ space obtained by integrating (\ref{probp}) 
with respect to $(\Omega_m,~b_0,~p)$. 
The target parameters are same as those in Figure 1.
Similarly, we set $z_*=0.13$ in (\ref{wz}). 
In this case, it is clear from Figure 2 that the 
degeneracy between $\bar w$ and $\nu$ is broken. 
Note that $\bar w=w_X(z_*)$. Thus the 
power spectrum analysis of the LRG sample gives the 
equation of state around $z=0.13$, almost independently from $\nu$. 
Conversely, the LRG sample is not very sensitive to probe $\nu$.


The minimum error in determining the equation of state
is $\Delta \bar w\simeq 0.1$, which is
given by integrating the probability function over the other 
parameters $(\Omega_m,~\nu,~b_0,~p)$.
The error can be reduced, when $\Omega_m$ is determined from other 
observation independently.
The result depends on the target parameters. Figure 3 shows
the $1-\sigma$ error $\Delta \bar w$ as function of $\Omega_m^{\rm tr}$.
The solid curves assume $\bar w^{\rm tr}=-1$, while
the dashed curves assume $\bar w^{\rm tr}=-0.8$.
The other target parameters are same as those in Figure 1.
The error is sensitive to the density parameter $\Omega_m^{\rm tr}$, 
however, is not sensitive to $\bar w^{\rm tr}$. The sensitivity to 
the equation of state deteriorates in higher matter 
density universes. 
This is because the fraction of the dark energy to the total 
energy decreases as the matter density becomes higher. 
As long as $\Omega_m^{\rm tr}\simlt 0.3$, however, 
the minimum error is $\Delta \bar w\simlt 0.1$. 

\subsection{A Luminosity Dependent Bias Model}

Here we consider a more realistic bias model.
Recent investigations with the 2df galaxy redshift survey
report that the clustering amplitude depends on galaxy 
type and luminosity (Norberg et~al. 2001, 2002).
The selection effect from an apparent limiting magnitude
can have significant influences on the redshift 
evolution of the bias. Denoting the luminosity dependent 
bias by $b(z,L)$, we define the luminosity-averaged bias as
\begin{eqnarray}
  \bar b(z)={\int_{L_{\rm min}(z)}^\infty\phi(L) b(z,L)dL
  \over \int_{L_{\rm min}(z)}^\infty\phi(L) dL},
\label{barb}
\end{eqnarray}
where $\phi(L)$ is the luminosity function. In modeling 
$b(z,L)$, we assume that $b(z,L)$ can be approximately 
written as $b(z,L)=\hat b(z) b(z=0,L)$.
Namely, the luminosity dependence of the bias does not depend
on the redshift. Here we also assume $\hat b(z)$ to be described 
by (cf. Mo \& White 1996)
\begin{eqnarray}
  \hat b(z)={1\over b_0}\biggl(1+{b_0-1\over D(z)}\biggr),
\label{halob}
\end{eqnarray}
which is normalized to yield $1$ at $z=0$, i.e.
$\hat b(0)=1$. For $b(z=0,L)$, 
we adopt the result by Norberg et~al. (2002),
who found that the luminosity dependence can be 
fitted by
\begin{eqnarray}
  b(z=0,L)\propto A+(1-A){L\over L_*},
\label{lumib}
\end{eqnarray}
where $A$ is constant. 
Norberg et~al. report $A=0.8$ for early-type galaxy (2002). 
Combining (\ref{halob}) and (\ref{lumib}),
we assume $b(z,L)$ can be written in the form 
\begin{eqnarray}
  b(z,L)=\biggl(1+{b_0-1\over D(z)}\biggr)
  \biggl(A+(1-A){L\over L_*}\biggr).
\end{eqnarray}
Then, expression (\ref{barb}) yields
\begin{eqnarray}
  \bar b(z)=\biggl(1+{b_0-1\over D(z)}\biggr)
  \biggl(A+(1-A){\Gamma(\alpha+2,x(z))\over\Gamma(\alpha+1,x(z)) }\biggr),
\label{barbz}
\end{eqnarray}
where $\Gamma(\beta,x)$ is the incomplete Gamma function
and $x(z)=L_{\rm lim}(z)/L_*$, and we assumed the 
Schechter luminosity function 
\begin{eqnarray}
  \phi(L)dL=\phi^* \biggl({L\over L_*}\biggr)^\alpha 
  \exp\biggr(-{L\over L_*}\biggr)d\biggl({L\over L_*}\biggr).
\end{eqnarray}
In the present paper, we adopt the fitting formula for 
the luminosity function in the reference by Madgwick et~al. 
(2002) for early-type galaxies ($\alpha=-0.54$ and 
$M_*-5\log_{10}h=-19.58$), 
and $m=19.5$ for the apparent limiting magnitude. 
Figure 4 shows $\bar b(z)$ as a function of $z$, where 
we set $b_0=1.2$ and $A=0.8$, and we have not considered 
the luminosity evolution and the k-correction for simplicity.


We consider the bias parameterized by (\ref{barbz}) with $b_0$ and $A$, 
instead of (\ref{bz}) with $b_0$ and $p$, and repeat the evaluation of 
the Fisher matrix. Here we fix the target parameters, $\Omega_m=0.28$, 
$\bar w=-1$ and $\nu=0$. 
Figure 5 shows the marginalized best statistical error 
$\Delta \bar w$, integrating over the parameters except for
$\bar w$ as function of the target parameters $b_0$ with $A$ 
fixed $A=0.6$ (short-dashed curve), $A=0.8$ (solid curve), $A=0.9$ 
(long-dashed curve) and $A=1$ (dot-dashed curve). 
Thus, when the bias evolution can be fitted 
in the form of (\ref{barbz}), $\Delta\bar w\simeq0.1$, which is
almost the same level as that in Figure 3 for $A\simlt0.9$. 
However, for the case $A=1$, the error $\Delta \bar w$ 
increases out to 0.15, depending on $b_0$.
This suggests that modeling of the redshift-evolution of 
the bias can be problematic to constrain the equation of 
state of dark energy using the LRGs.

\section{OPTIMIZED SAMPLE}
In this section, we address the problem of determining the
optimized sample in order to best constrain the equation
of state. This problem might be only of theoretical interest,
but having such information could be useful in planning a 
survey of galaxies or clusters of galaxies. The same problem 
has been considered with regard to supernova data and lensing
systems by several authors (Huterer \& Turner 2001, Spergel \& 
Starkman 2002, Yamamoto et~al. 2001).
To address the problem stated above, we introduce
the Fisher matrix element per object as follows:
When a survey volume is small enough so that the 
light-cone (redshift evolution) effect is negligible, 
(\ref{kint}) is written
\begin{eqnarray}
  {\bfF}_{ij}\simeq {1\over 4\pi^2} 
  \int _{k_{\rm min}}^{k_{\rm max}} 
  {\Delta V\bar n^2\over (1+\bar n P_0(k,s))^2} 
  {\partial P_0(k,s)\over \partial \theta_i}
  {\partial P_0(k,s)\over \partial \theta_j}
  k^3 d\ln k,
\end{eqnarray}
where we assumed $\bar n$ to be constant, and $\Delta V$ 
is the survey volume.
Then ${\bfF}_{ij}/\Delta V$ can be regarded as
the Fisher matrix element per unit volume, and similarly
${\bfF}_{ij}/(\Delta V \bar n)$ is regarded as the Fisher 
matrix element per object. In a strict sense we must choose
$k_{\rm min}$ and $k_{\rm max}$ depending on $\Delta V$
and $\bar n$. However, we here fix 
$k_{\rm min}=0.01~h{\rm Mpc}^{-1}$ and 
$k_{\rm max}=1~h{\rm Mpc}^{-1}$.
Then we investigate ${\bfF}_{ij}/(\Delta V \bar n)$ as a function 
of $z$ and $\bar n$. Here note that $s$ is the function of the 
redshift, i.e. $s=s(z)$.

In this section, for simplicity, we consider the model
in which the equation of state is constant 
$w_X(z)=\bar w$, which is equivalent to the assumption
$f_X(z)=(1+z)^{3(1+\bar w)}$. Regarding the Fisher 
matrix as the $2\times 2$ matrix corresponding to 
the elements $\bar w$ and $b_0$, 
we consider the marginalized Fisher matrix element 
integrating the probability function over $b_0$:
\begin{equation}
\tilde{\bfF}_{\bar w\bar w}
={\bfF}_{\bar w\bar w}-{{\bfF}_{\bar w b_0}^2\over {\bfF}_{b_0 b_0}}.
\end{equation}
Figure 6 shows $[\Delta V \bar n/\tilde{\bfF}_{\bar w\bar w}]^{1/2}$
as a function of $z$ with $\bar n$ fixed as
$\bar n=10^{-2}~h^3{\rm Mpc}^{-3}$~(dashed curve), 
$10^{-4}~h^3{\rm Mpc}^{-3}$~(solid curve), and 
$10^{-6}~h^3{\rm Mpc}^{-3}$~(long dashed curve). 
Figure 6 indicates that objects in the range
$0.4\simlt z\simlt 1$ are most efficient for measuring $\bar w$, 
because $[\Delta V \bar n/\tilde{\bfF}_{\bar w\bar w}]^{1/2}$
is regarded as the error in determining $\bar w$ per object 
at the redshift $z$. 
Note that the error using the total objects $\Delta \bar w$ scales as $1/
\sqrt{N}=1/\sqrt{\bar n \Delta V}$.
Figure 7 shows $[\Delta V \bar n/\tilde{\bfF}_{\bar w\bar w}]^{1/2}$
as a function of $\bar n$ with the redshift fixed at $z=0.6$. 
This figure indicates that the number density around $\bar n =10^{-4}
~h^3{\rm Mpc}^{-3}$ is most efficient for measuring $\bar w$.
For very sparse samples, like quasars,  which have a typical number 
density $\bar n =10^{-6}~h^3{\rm Mpc}^{-3}$, the errors in 
measuring parameters using the power spectrum is large due to the shot-noise.
On the other hand, for a sample with high number density, 
the shot-noise is of minor importance, but the efficiency in constraining 
$\bar w$ per object decreases as well. Thus the objects
in the range of redshift $0.4\simlt z\simlt 1$ 
with the number density $\bar n\simeq 10^{-4}~h^3{\rm Mpc}^{-3}$ 
are the optimized sample.

\section{CONCLUSIONS}

In summary we derived a rigorous optimal weighting scheme 
for power spectrum analysis, which is useful for samples
in which the light-cone effect and the redshift-space 
distortions are substantial. Our result is a simple 
extension of the work by FKP, and we obtained a 
generalized optimal weight factor which minimizes the 
errors of the power spectrum estimator.
As an application of our formula, we 
investigated the capability of the LRG sample
in SDSS to constrain the equation of 
state parameters by evaluating Fisher matrix elements.
Even if the transfer function for the matter power spectrum 
does not depend on $w_X$, the power spectrum analysis
of the redshift-space sample can constrain $w_X$
due to the geometric distortion.
To incorporate uncertainties of redshift evolution of
the clustering bias, we considered the marginalized 
probability function by integrating over the parameters 
of the bias models. 
This analysis shows
that the LRG sample in SDSS has a serviceable potential for
constraining the equation of state around $z=0.13$ 
with $1$ sigma errors at the $10$ \% level,
if other fundamental 
parameters are well determined in an independent fashion.
We also showed that this conclusion is not altered
in the case of a bias model incorporating the redshift-evolution
due to selection effect depending on luminosity. 
However, even in our realistic treatment of the bias, 
we made simplifications:
Uncertainties including the stochasticity and 
the nonlinearity in modeling the bias are not 
considered in our investigation. Then, tests on the
bias properties will be required for more definite conclusions.

In the present paper, we assumed $10^4$ deg${}^2$ as 
the complete SDSS survey area. When the planed survey 
area are not achieved, the capability to constrain
the equation of state reduces. 
As the Fisher matrix element is in proportion to
the survey area $\Delta \Omega$, then the 
statistical error $\Delta \bar w$ increases 
in proportion to $\sqrt{\Delta \Omega}$.
For example, when we assume $5\times 10^3$ deg${}^2$ 
and $7.5\times 10^3$ deg${}^2$ 
as the final SDSS survey area, $\Delta \bar w$
increases by $30$ \% and $15$ \%, respectively, 
as long as the inhomogeneity
of the incomplete survey area does not cause
additional systematic errors. 

In section 4, we considered the optimized sample to constrain
$w_X$ using power spectrum analysis. We found 
that it is most advantageous to have the sample with the 
comoving number density $\bar n\simeq 10^{-4}~h^3{\rm Mpc}^{-3}$
in the range of redshift $0.4\simlt z \simlt 1$.
For such a sample,
the efficiency per object to constrain $w_X$ is optimized.
Information from anisotropic power spectrum would improve the 
capability of constraining the parameters, as demonstrated in 
the 2df QSO sample (Outram et~al. 2001).


\acknowledgements{
This work was supported by fellowships for Japan Scholar 
and Researcher Abroad from Japanese Ministry of Education, Culture, Sports, 
Science and Technology. The author thanks Prof. S. D. M. White for
important comments, which helped
improve the paper. Parts of section 3 are based on the 
valuable communications with him. The author is also grateful to 
the people at Max-Planck-Institute for Astrophysics (MPA) 
for their hospitality during his stay. 
He thanks P. A. Popowski, N. Sugiyama, S. Zaroubi, 
J. Brinchmann and T. Matsubara for useful discussions 
and comments.
He is also grateful to B. M. Sch\"afer and K. Basu 
for reading manuscript and useful comments.}

\clearpage
\begin{figure}
\begin{center}
    \leavevmode
    \epsfxsize=16cm
    \epsfbox[20 150 600 720]{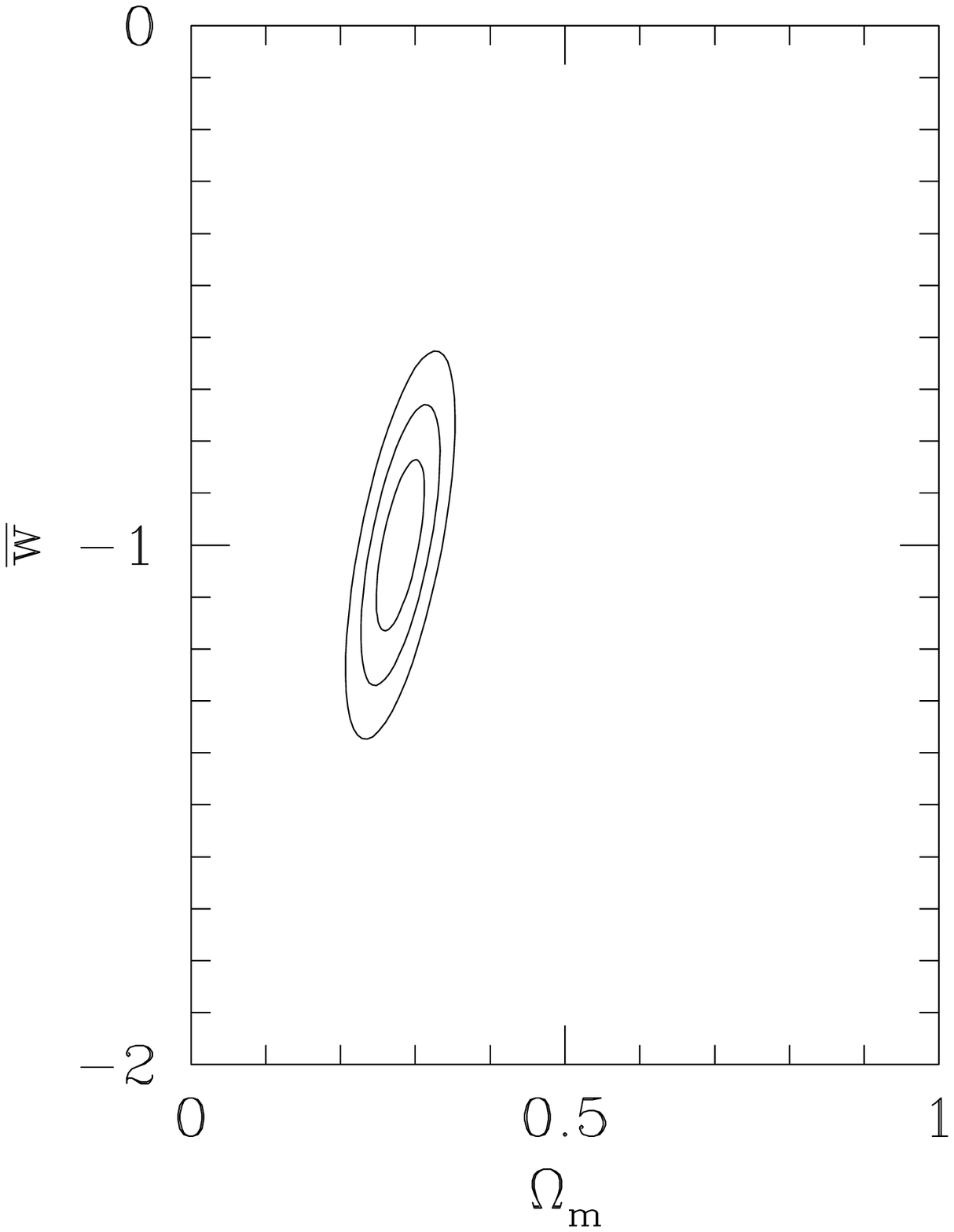}
\end{center}
\caption{Contour of the marginalized probability 
function in $\Omega_m$-$\bar w$ space. The curves
show $1-\sigma$, $2-\sigma$ and $3-\sigma$ contours.
The target parameters are 
$\Omega_m=0.28$, $\bar w=-1$, $\nu=0$, $b_0=1.5$ and $p=1$. 
In (\ref{wz}) we set $z_*=0.13$. The survey area 
$10^4~{\rm deg}^2$ is assumed.
}
\label{fiburea}
\end{figure}
\begin{figure}
\begin{center}
    \leavevmode
    \epsfxsize=16cm
    \epsfbox[20 150 600 720]{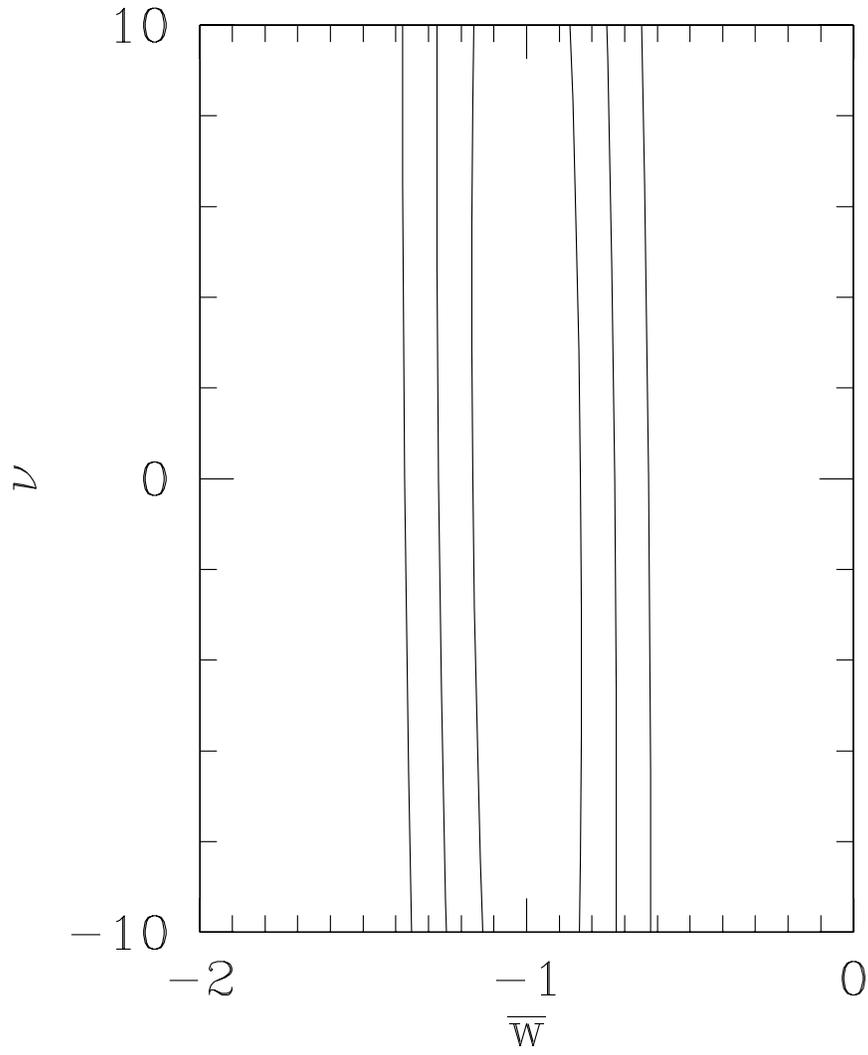}
\end{center}
\caption{Contour of the marginalized probability 
function in $\bar w$-$\nu$ space. The meaning of
the curves and the parameters are same as those in Figure 1.
}
\label{figureb}
\end{figure}
\begin{figure}
\begin{center}
    \leavevmode
    \epsfxsize=16cm
    \epsfbox[20 150 600 720]{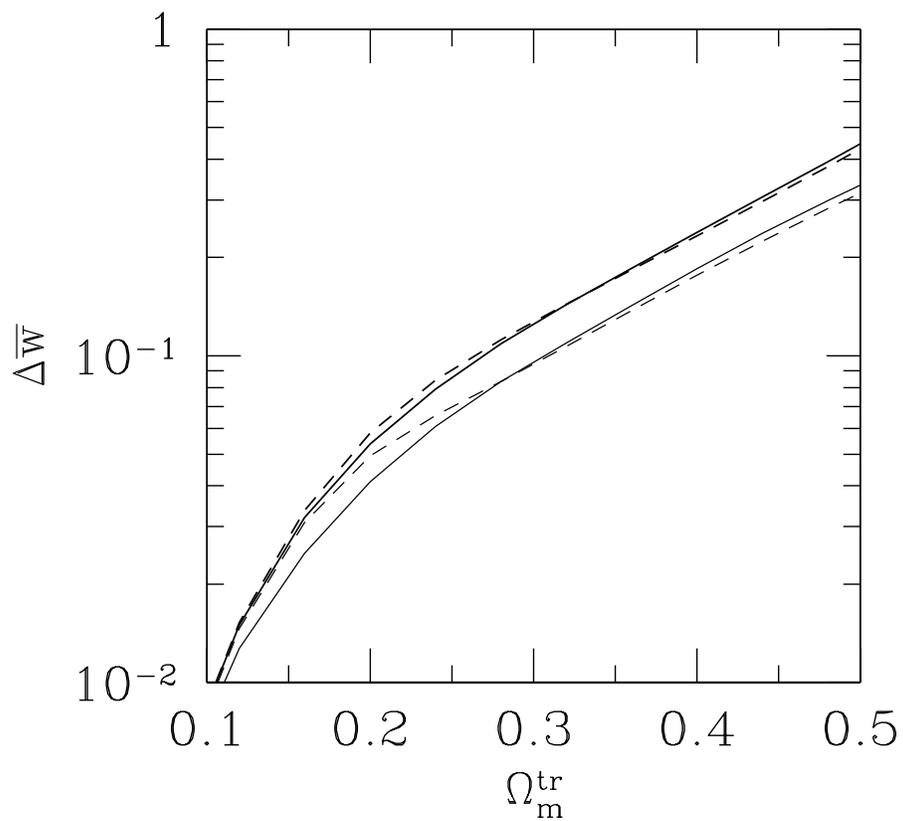}
\end{center}
\caption{
The statistical $1-\sigma$ error $\Delta \bar w$ as a function 
of the target parameter $\Omega_m^{\rm tr}$.  
The solid and dashed curves adopt $\bar w^{\rm tr}=-1$ 
and $\bar w^{\rm tr}=-0.8$, respectively. 
For each pair of curves, 
the lower curve assumes that $\Omega_m$ is 
determined by other independent method, while the upper curve
assumes that $\Omega_m$ is determined simultaneously by the
power spectrum analysis. 
The other target parameters are same as those in Figure 1. 
}
\label{figurec}
\end{figure}
\begin{figure}
\begin{center}
    \leavevmode
    \epsfxsize=16cm
    \epsfbox[20 150 600 720]{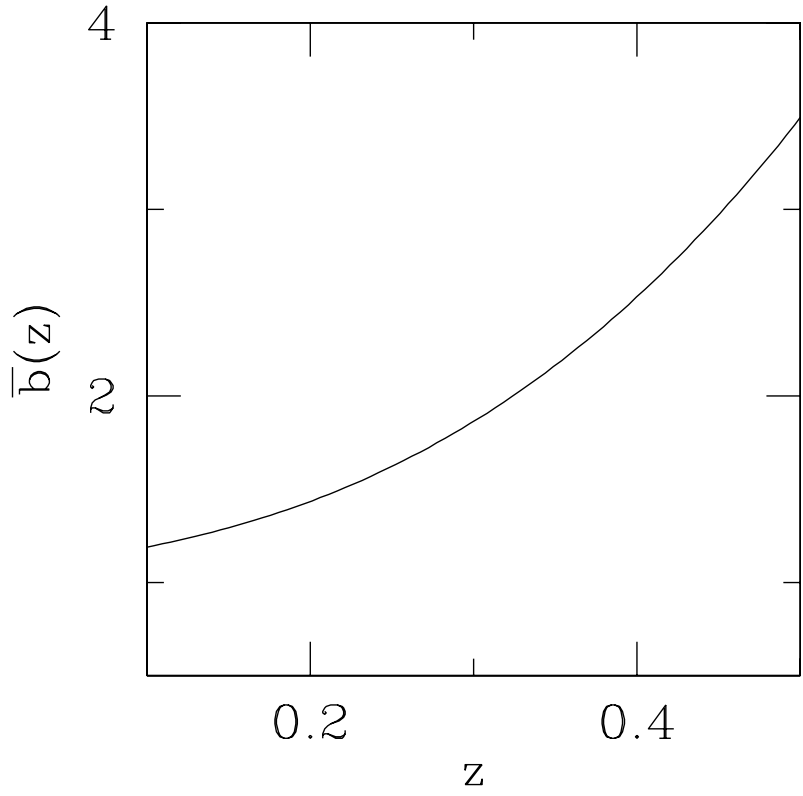}
\end{center}
\caption{$\bar b(z)$ as a function of the redshift. 
We adopted $b_0=1.2$ and $A=0.8$.}
\label{figureg}
\end{figure}
\begin{figure}
\begin{center}
    \leavevmode
    \epsfxsize=16cm
    \epsfbox[20 150 600 720]{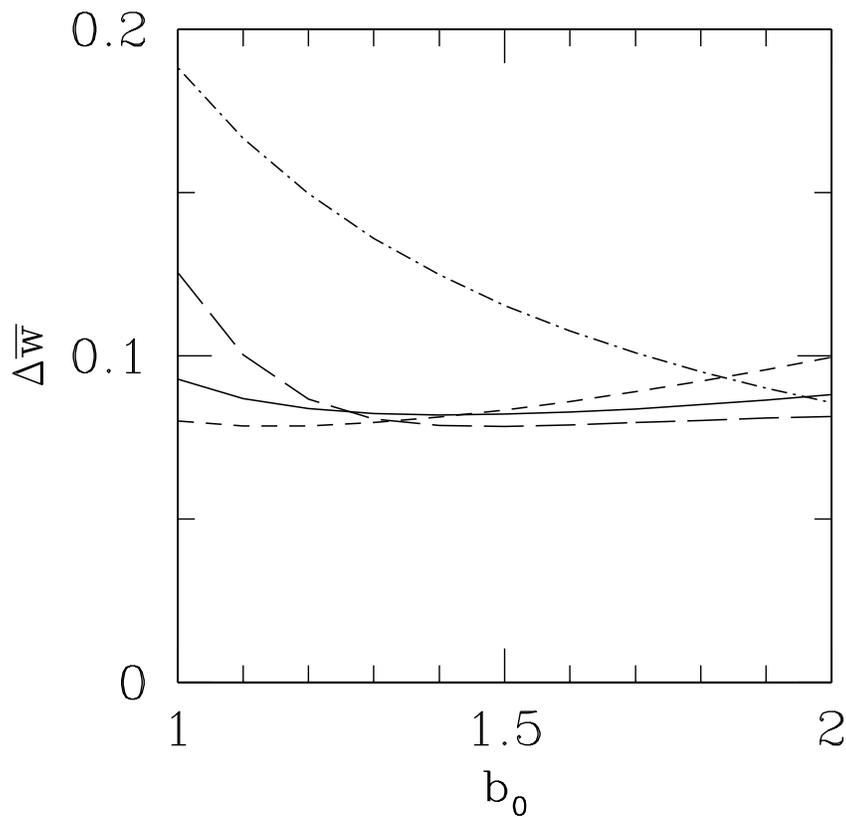}
\end{center}
\caption{ 
The statistical error $\Delta \bar w$ as a function of the target 
parameter $b_0$ with fixing
$A=0.6$ (short-dashed curve), $A=0.8$ (solid curve), $A=0.9$ (long-dashed 
curve) and $A=1$ (dot-dashed curve) in the bias model in section 3.3.
The other target parameters are same as those in Figure 1. 
Here we assume that $\Omega_m$ is determined simultaneously 
by the power spectrum analysis. }
\label{figureh}
\end{figure}

\begin{figure}
\begin{center}
    \leavevmode
    \epsfxsize=16cm
    \epsfbox[20 150 600 720]{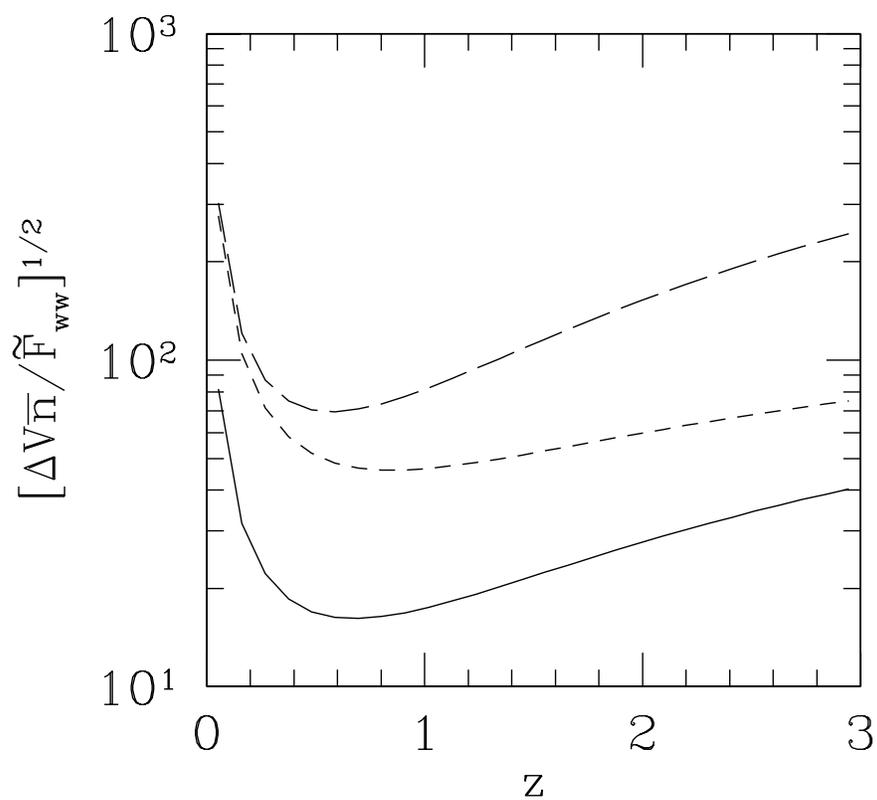}
\end{center}
\caption{
$[\Delta V \bar n/\tilde{\bfF}_{\bar w\bar w}]^{1/2}$
as function of $z$ with $\bar n$ fixed.
The dashed curve, the solid curve, and the
long dashed curve, assume  
$\bar n=10^{-2}~h^3{\rm Mpc}^{-3}$,
$10^{-4}~h^3{\rm Mpc}^{-3}$, and 
$10^{-6}~h^3{\rm Mpc}^{-3}$, respectively. 
The target parameters are same as those in Figure 1.
}
\label{figured}
\end{figure}
\begin{figure}
\begin{center}
    \leavevmode
    \epsfxsize=16cm
    \epsfbox[20 150 600 720]{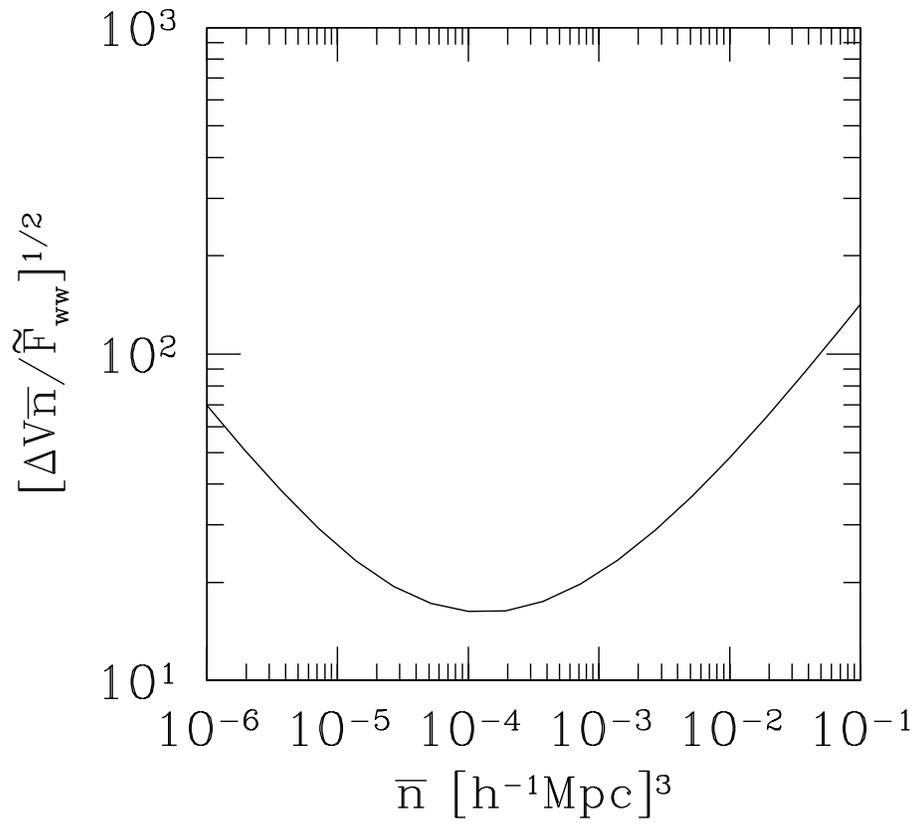}
\end{center}
\caption{$[\Delta V \bar n/\tilde{\bfF}_{\bar w\bar w}]^{1/2}$
as a function of $\bar n$. Here the redshift is fixed at $z=0.6$.
The target parameters are same as those in Figure 1.}
\label{figuree}
\end{figure}

\end{document}